\documentstyle[preprint,aps]{revtex}

\begin{document}
\draft
\tighten

\preprint{\vbox{\hbox{IMPERIAL/TP/94-95/11}
\hbox{TUTP-95-2}
\hbox{NI94039}
\hbox{hep-ph/9501266}}}

\title{Phase equilibration in bubble collisions}

\author{T.W.B. Kibble}
\address{Blackett Laboratory, Imperial College, London SW7 2BZ,
United Kingdom and\\
Isaac Newton Institute for Mathematical Sciences, Cambridge CB3 0EH,
United Kingdom}
\author{Alexander Vilenkin}
\address{Institute of Cosmology,
Department of Physics and Astronomy, \\
Tufts University, Medford MA 02155, USA and\\
Isaac Newton Institute for Mathematical Sciences, Cambridge CB3 0EH,
United Kingdom}

\date{}

\maketitle

 \begin{abstract}
In the context of an Abelian gauge symmetry, spontaneously broken at a
first-order transition, we discuss the evolution of the phase
difference between the Higgs fields in colliding bubbles.  We show that
the effect of dissipation, represented by a finite plasma conductivity,
is to cause the phases to equlibrate on a time-scale, determined by the
conductivity, which can be much smaller than the bubble radii at the
time of collision.  Currents induced during the phase equilibration
generate a magnetic flux, which is determined by the initial phase
difference.  In a three-bubble collision, the fluxes produced by each
pair of bubbles combine, and a vortex can be formed.  We find that,
under most conditions, the probability of trapping magnetic flux to
form a vortex is correctly given by the ``geodesic rule''.
 \end{abstract}


\narrowtext

\section{Introduction}

Phase transitions in the early universe, as in condensed-matter
systems, may lead to the formation of defects, such as cosmic strings
\cite{reviews}.  To determine the observational implications of defects
one must follow their evolution from an initial state soon after the
phase transition up to the time when they begin to have visible
effects.   There has recently been renewed interest in the problem of
estimating the initial defect density, which provides the starting
point for such studies.

In this paper, we are concerned with the initial number density of
strings formed at a first-order phase transition proceeding by bubble
nucleation.  For simplicity, we consider an Abelian theory with a
scalar Higgs field $\Phi$, described by the Lagrangian
 \begin{equation}
{\cal L}=D_\mu\Phi^*D^\mu\Phi
-\case{1}{4}F_{\mu\nu}F^{\mu\nu}-V(\Phi),
 \end{equation}
where $D_\mu\Phi=\partial_\mu\Phi+ieA_\mu\Phi$,
$F_{\mu\nu}=\partial_\mu A_\nu-\partial_\nu A_\mu$, and the potential
$V$ is a function of $|\Phi|$ with a local minimum at $|\Phi|=0$ and a
global minimum at $|\Phi|=\eta/\surd2$.

Nothing happens at the theoretical critical temperature $T_{\rm c}$
(where the two minima of the high-temperature effective potential are
degenerate).  Once the temperature has fallen well below $T_{\rm c}$,
tunnelling can occur from the false vacuum state at $\Phi=0$ to the
true vacuum, where $\Phi\approx\eta e^{i\theta}/\surd2$.  This is an
example of spontaneous symmetry breaking: within each bubble there is a
random choice of the phase angle $\theta$.

The bubbles nucleate at random points in space and then expand.  The
bubble walls accelerate, often nearly reaching the speed of light,
until they meet and fill the whole of space.  The nucleation rate per
unit space-time volume is determined by the tunnelling probability,
which may be calculated from the ``bounce'' solution of the Euclidean
field theory \cite{Coleman}.  It is likely to be a rapidly rising
function of time, but for purposes of illustration let us assume that
it is zero up to some given time and thereafter a constant $\gamma$.
We also suppose that the bubble walls expand at speed $v$, with $\gamma
v^3\gg H^4$ (where $H$ is the Hubble parameter).  Then the final mean
number density of bubbles will be approximately $1/\xi^3$, where
$\xi=(v/\gamma)^{1/4}$.  (These assumptions are in no way critical to
our discussion.)

The traditional picture of string formation is as follows
\cite{Kib76}.   Within each bubble the phase $\theta$ is essentially
constant, but phases in different bubbles are uncorrelated.  When two
bubbles with phases $\theta_1$ and $\theta_2$ meet, the sharp
discontinuity in the phase in smoothed out to become a smooth
variation.  For energetic reasons, the system will tend to choose the
shorter of the two paths between $\theta_1$ and $\theta_2$.  For
example, if $\theta_1=0$ and $0<\theta_2<\pi$, the phase will increase
from $0$ to $\theta_2$; but if $\pi<\theta_2<2\pi$, it will instead
decrease from $0$ to $\theta_2-2\pi$.  This is often called the
``geodesic rule''.  It has an obvious extension to higher symmetry
groups.

When three bubbles meet, a string or vortex may be trapped in the
region between them.  Whether this happens or not is determined by the
net phase change in going from $\theta_1$ to $\theta_2$, to $\theta_3$
and back to $\theta_1$.  It is easy to see that if the three angles are
independent random variables, uniformly distributed between $0$ and
$2\pi$, and if the geodesic rule is correct, then the probability of
trapping a string is $1/4$.  Thus the initial string density (length
per unit volume) is of order $1/4\xi^2$.

This is obviously only a crude estimate.  Several of the assumptions
we have made are questionable.  Some of the objections are relatively
minor.  It is clearly not precisely true that the phase within each
bubble is constant.  There must be random thermal and quantum
fluctuations.  So the geodesic rule will sometimes break down.  When
two bubbles meet, there may be local random fluctuations that will
cause the phase to go the ``wrong'' way round in joining $\theta_1$ and
$\theta_2$.  But the effect of this would merely be a slight increase
in the probability of trapping a string.  Indeed, there could also be a
nonzero probability that the net phase change would be $\pm4\pi$, thus
trapping a string of winding number $\pm2$.

As the bubbles grow, closed loops within which strings may be trapped
will often be formed of more than three bubbles.  However, as the
bubbles expand further into the trapped region, it will generally be
split up into several smaller regions.  Eventually, the remaining gaps
will always be finally closed by the junction of three bubbles.

There is a particular problem in cases where the velocity of the
bubble walls is low.  When $v\approx c$ we can be assured that no
bubble collision can be causally affected by any previous collision
(because any two points on the expanding light cone are spacelike
separated).  But when $v\ll c$, phase equilibration within a composite
bubble might have been completed before the two components encounter a
third bubble, thus reducing the chances of trapping a string.  Of
course, some strings would still be trapped, because the three
collisions would sometimes occur nearly simultaneously, but the string
density would certainly be less.  One very interesting open question
(to which we hope to return) is whether in that event the relative
proportion of small loops to long strings would be changed.  At first
sight one might think that if strings are rare then most of them would
form small loops \cite{Vachaspati}.  If so, that would have a dramatic
impact on the resulting cosmology.

Underlying all of these points is a more serious concern: does it make
sense to talk of phase differences between bubbles in a gauge theory?
Indeed by a gauge transformation, the phase difference can be set equal
to any value we please.  It has been argued by Rudaz and Srivastava
\cite{RudSri} that string formation in gauge theories might be strongly
suppressed, although a recent analysis by Hindmarsh {\it et al\/}
\cite{HinBraDav} suggests that it is not.

It is clearly important to understand the process of string formation
in gauge theories.  We begin by studying what happens when two bubbles
meet.  An important feature of the process of phase equilibration (not
included in previous analyses) is the role of dissipation.  We include
an ohmic dissipation term in our equations.  However, it will be useful
first to consider the problem in the absence of dissipation, especially
to clarify some issues relating to gauge choice and boundary conditions.

There are two ways of dealing with the problem of gauge invariance.
One is to use only gauge-invariant quantities, such as the covariant
derivative $D_\mu\theta = \partial_\mu \theta +eA_\mu$.
This approach provides valuable insights, but it is difficult to
apply to the string-formation problem because we have to consider
non-local variables such as the line integral of $D_\mu \theta$.
The other is to choose a gauge, but that too has its problems.  We
need a {\it complete\/} gauge choice, not merely a class of gauges like
the Lorentz gauge within which some gauge freedom still exists.

There are several gauges we might choose:

I.  The Coulomb gauge.  This works well provided we are considering
only a limited region, outside of which there are no relevant charges.
It is problematic in other cases, because boundary conditions are
needed to solve the Poisson equation for the scalar potential
$\phi\equiv A^0$.  In practice, this gauge is not as convenient as the
others, so we shall not consider it further.

II.  The unitary gauge, defined by setting the phase angle $\theta$ to
zero.  This is very convenient within any simply connected region where
$\Phi\ne0$.  Unfortunately, it does not work if the region is
non-simply connected or contains zeroes of $\Phi$, {\it i.e.}, when
strings are present.  It will be useful in discussing the collision of
two bubbles, but is less easy to use to study three-bubble collisions.

III.  The axial gauge defined by $A^z=0$, together with a choice of
gauge on the surface $z=0$.  This is particularly well suited to the
problem of two colliding bubbles, where it matches the symmetry of the
problem and provides a very neat solution, as we shall see in the next
section.  However, it is not very convenient for more general problems.

IV.  The temporal gauge, given by $A^0=0$.  This is universally
applicable, and very simply related to the unitary gauge.  To define it
completely, we need a gauge choice at some initial time, but in our
case this is straightforward: we assume that no electromagnetic fields
are present before bubble nucleation starts, so that we can simply set
${\bf A}={\bf 0}$ at that time.  The gauge is then defined for all
times.  However, this gauge does introduce somewhat artificial spatial
discontinuities in various functions.

It will be useful first to examin the collision of two bubbles with no
damping of the phase oscillations.  Although this is a rather
artificial problem, the results will help to motivate some of the more
general discussion later.   We begin by defining the gauge-invariant
phase difference which plays an important role in determining the
outcome of the collision.

\section{Gauge-invariant phase difference}

It is convenient to write the Higgs field in polar form:
 \begin{equation}
\Phi={1\over\surd2}Xe^{i\theta}.
 \end{equation}
Then the equation of motion for $\theta$ is simply the continuity
equation $\partial_\mu j^\mu=0$ for the current
 \begin{equation}
j_\mu = -eX^2D_\mu\theta \equiv
-eX^2(\partial_\mu\theta+eA_\mu).
\label{Cur}
 \end{equation}
The phase gradient $D_\mu \theta$ is gauge-invariant and can be used
to define a gauge-invariant phase difference between the two bubbles,
 \begin{equation}
\Delta \theta = \int_A^B dx^k D_k \theta ,
\label{GIPD}
 \end{equation}
where $k=1,2,3$ and points $A$ and $B$ are taken in bubble interiors.
While the bubbles are well-separated, $D_\mu \theta$ is appreciably
different from zero only in the region where $X \approx 0$, so that
$j_\mu \approx 0$ and $F_{\mu\nu} \approx 0$ (assuming that
$F_{\mu\nu}=0$ prior to bubble nucleation).  Then $\Delta \theta$ is
independent of the choice of path between $A$ and $B$.  In general of
course, $\Delta\theta$ is path-dependent, and to make it well-defined,
we shall choose the integration path to be along the straight line
passing through the bubble centers.

To analyse what happens when the bubbles collide,
we shall assume that the {\it radial\/} mode is strongly damped, so
that $X$ settles
rapidly to its equilibrium value $\eta$.  The equation for $\theta$ is
then the Klein-Gordon equation
 \begin{equation}
(\partial^2 +m_A^2 ) D_\mu \theta = 0
\label{KleinGordon}
 \end{equation}
with the gauge boson mass $m_A =e\eta$.

Let us first consider the central region of the collision, where the
bubble walls can be well approximated as parallel approaching planes,
so that the problem becomes one-dimensional.  The initial
configuration, just after the bubbles have collided, say at $z=0$ and
$t=0$, will then have $D_x \theta =D_y \theta =0$ and $D_z \theta$,
$D_t \theta$ concentrated near $z=0$ within some distance $\delta$.  If
the energy of the colliding walls is ``instantly'' dissipated, then
$\delta$ is comparable to the Lorentz-contracted wall thickness.
(However, if the motion of the walls is relativistic, they are likely
to overshoot and go through several oscillations before loosing their
energy \cite{Hawking}.  In this case the collision region will be much
wider and the phase equilibration process may be more complicated).

It is not difficult to understand that in the course of the following
evolution the ``wave packets'' of $D_z \theta$ and $D_t \theta$ will
spread in the $z$-direction and oscillate at the frequency $\omega =
m_A$.  The speed of spreading, $u$, will be determined by the initial
width of the packets: $u \sim 1/m_A \delta$ for $\delta \gg m_A^{-1}$
and $u \sim 1$ for $\delta \lesssim m_A^{-1}$.  (This qualitative
picture is confirmed by explicit solutions of Eq.(\ref{KleinGordon})
with Gaussian ``packets'').

An equation for the phase difference
 \begin{equation}
\Delta\theta = \int_{-z_0}^{z_0} dz D_z \theta
 \end{equation}
can be obtained by integrating Eq.(\ref{KleinGordon}) over $z$,
 \begin{equation}
\left({d^2 \over{dt^2}} + m_A^2 \right) \Delta\theta = 0.
\label{HarmOsc}
 \end{equation}
The boundary terms vanish as long as $z_0 \gg ut$.  The solution of
(\ref{HarmOsc}) is $\Delta\theta \propto \cos (m_At)$.  We note the
``acausal'' behavior of $\Delta\theta$: it varies on a microscopic
timescale $\sim m_A^{-1}$ which can be much smaller than the radii of
the colliding bubbles.

To extend the analysis beyond the central collision region, we shall
now fix the gauge and solve the field equations for $\theta$ and
$A_\mu$.  Different gauges provide useful, and complementary, insights.

\section{Collisions without dissipation}

We consider the simplest possible case, where the bubble walls are
assumed to move essentially with the speed of light.   Without loss of
generality, we can choose a frame of reference in which the two bubbles
nucleate simultaneously, say at the points $(0,0,0,\pm R)$.  The
bubbles first collide at $(R,0,0,0)$, when the radii are $R$.  The
problem has a high degree of symmetry: it is invariant under the
3-dimensional Lorentz group SO(1,2) in the $(t,x,y)$ subspace
\cite{Coleman,Hawking}.   The bubble collision occurs along the surface
$z=0$, $t^2-x^2-y^2=R^2$; for any point on that surface there is a
frame of reference in which that is the point of first contact.

The obvious gauge to use here is the axial gauge which picks out the
one distinguished direction, the vector joining the nucleation
centres.  The symmetry of the problem shows that in this gauge we must
have
 \begin{equation}
\theta(x)=\theta(\tau,z), \qquad
A^\alpha(x)=x^\alpha a(\tau,z),
 \end{equation}
where $\alpha=(0,1,2)$ and
 \begin{equation}
\tau^2 = x^\alpha x_\alpha = t^2-x^2-y^2.
 \end{equation}
The $z$ component of the Maxwell field equation, with the current
(\ref{Cur}), yields (with a suitable choice of arbitrary constant),
 \begin{equation}
\theta(\tau,z) = {1\over e\eta^2}(3a+\tau\partial_\tau a).
\label{ThetaA}
 \end{equation}
Thus there is just one independent unkown function.

The remaining Maxwell equations, together with the continuity equation
for the current, show at once that both $\theta$ and $A^\alpha$ obey
the Klein-Gordon equation with mass $e\eta$, whence the functions
$\theta$ and $a$ satisfy the equations
 \begin{eqnarray}
\partial_\tau^2\theta+{2\over\tau}\partial_\tau\theta
-\partial_z^2\theta+e^2\eta^2\theta&=0,\nonumber\\
\partial_\tau^2 a+{4\over\tau}\partial_\tau a-\partial_z^2 a
+e^2\eta^2 a&=0.
 \end{eqnarray}

For the initial conditions we assume that as we go from one bubble
into the other, $\theta$ changes rapidly, say from $-\theta_0$ to
$\theta_0$, while $\dot\theta=0$.  Thus
 \begin{equation}
\theta|_{\tau=R}=\theta_0\epsilon(z),\qquad
\partial_\tau\theta|_{\tau=R}=0.
\label{ThetaInit}
 \end{equation}
We also require that $A^\alpha(x)=0$ up to and on the boundary of the
bubble overlap region.  Using (\ref{ThetaInit}) and (\ref{ThetaA}),
this yields
 \begin{equation}
a|_{\tau=R}=0,\qquad \partial_\tau a|_{\tau=R}
={e\eta^2\over R}\theta_0\epsilon(z).
 \end{equation}

It is straightforward to solve these equations.  The solution for
$\theta_{\rm a}$ (the subscript indicates the gauge choice) is
 \begin{equation}
\theta_{\rm a} = {\theta_0 R\over\pi\tau}\int{dk\over k}\sin kz
\left(\cos\omega(\tau-R)+{1\over\omega R}\sin\omega(\tau-R)\right),
 \end{equation}
where
 \begin{equation}
\omega^2 = {\bf k}^2+e^2\eta^2.
 \end{equation}
(It is also possible to write this explicitly in terms of the Bessel
function of order zero, $J_0(e\eta\sqrt{(\tau-R)^2-z^2})$.)

The gauge-invariant phase difference $\Delta\theta$ can now be deduced
from the asymptotic behavior of $\theta_{\rm a} (t,x,y,z)$ at $z \to
\pm \infty$,
 \begin{equation}
\Delta\theta (t)=\theta_{\rm a}(t,0,0,+\infty )
-\theta_{\rm a}(t,0,0,-\infty ).
 \end{equation}
This gives
 \begin{equation}
\Delta\theta = {2R\over{t}}\theta_0
\left(\cos e\eta(t-R)+{1\over e\eta R}\sin e\eta(t-R)\right).
\label{ThetaAsy}
 \end{equation}
Thus phase equilibration {\it does\/} occur in this gauge with a time
scale determined, as one might expect, by the size $R$ of the colliding
bubbles, with superimposed oscillations with frequency given by the
gauge-field mass.

Similarly, $a_{\rm a}$ is given by
 \begin{equation}
a_{\rm a} = {\theta_0 e\eta^2\over\pi\tau^3}\int{dk\over k}\sin kz
\left[-{\tau-R\over\omega^2 R}\cos\omega(\tau-R)
+\left({\tau\over\omega}+{1\over\omega^3R}\right)
\sin\omega(\tau-R)\right].
 \end{equation}

It is of course straightforward to transform to other gauges.  The
unitary-gauge fields are given by
$A_{\rm u}^\mu=A_{\rm a}^\mu+(1/e)\partial^\mu\theta_{\rm a}$.
These fields still exhibit the full symmetry of the problem; one finds
 \begin{equation}
A_{\rm u}^\alpha=x^\alpha b(\tau,z), \qquad
b(\tau,z) = {1\over e^2\eta^2}\partial_z^2 a(\tau,z),
\label{Auni}
 \end{equation}
and
 \begin{equation}
A_{\rm u}^z=-{1\over e}\partial_z\theta_{\rm a}(\tau,z).
 \end{equation}
The unitary-gauge fields again obey the Klein-Gordon equation together
with the initial conditions
 \begin{equation}
A_{\rm u}^\alpha|_{\tau=R}=0, \qquad
\partial_\beta A_{\rm u}^\alpha|_{\tau=R}=
{x^\alpha x_\beta\over R^2}{2\theta_0\over e}\delta'(z),
 \end{equation}
 \begin{equation}
A_{\rm u}^z|_{\tau=R}=-{2\theta_0\over e}\delta(z), \qquad
\partial_\alpha A_{\rm u}^z|_{\tau=R}=0.
 \end{equation}

Finally, we can transform to the temporal gauge which, as we shall see, has
some rather odd features.  We find $\theta_{\rm t}$ by integrating the
equation
 \begin{equation}
\dot\theta_{\rm t}=eA_{\rm u}^0.
 \end{equation}
We need an initial condition.  Assuming as before that outside the
bubble overlap region $\theta=\theta_0\epsilon(z)$, and using
(\ref{Auni}), we find
 \begin{equation}
\theta_{\rm t}=\theta_0\epsilon(z)+e\int_R^\tau \tau b\,d\tau
=\theta_{\rm a}+e\int_R^\tau \tau a\,d\tau.
 \end{equation}
It is interesting to note that in spite of our non-covariant choice of
gauge, $\theta_{\rm t}$ is still a function only of $\tau$ and $z$.
It satisfes exactly the same initial conditions (\ref{ThetaInit}) as
$\theta_{\rm a}$.  However it does {\it not\/} satisfy the Klein-Gordon
equation.  If we integrate the $\mu=0$ Maxwell equation (the Gauss law)
using the given initial conditions, we find
 \begin{equation}
\partial_k A_{\rm t}^k = e\eta^2[\theta_{\rm t} - \theta_0\epsilon(z)].
 \end{equation}
Thus $\theta_{\rm t}$ actually obeys the equation
 \begin{equation}
\partial^2\theta_{\rm t} + e^2\eta^2[\theta_{\rm t}
 - \theta_0\epsilon(z)] = 0.
 \end{equation}

The symmetry of the problem also shows up, even more remarkably
perhaps, in the fact that not only does $A_{\rm t}^0=0$ but in fact
$A_{\rm t}^\alpha=0$.  The only non-zero component of the gauge
potential is
 \begin{equation}
A_{\rm t}^z=\int_R^\tau \tau\partial_z a\,d\tau.
 \end{equation}
Both $A_{\rm t}^z$ and its derivatives vanish on the initial surface
$\tau=R$.  It is non-zero only because it too does {\it not\/} satisfy
the Klein-Gordon equation; rather
 \begin{equation}
\partial^2 A_{\rm t}^z+e^2\eta^2\left(A_{\rm t}^z
-{2\theta_0\over e}\delta(z)\right) = 0.
 \end{equation}

One effect of the extra terms in the field equations is that
at late times, $\theta_{\rm t}$, unlike $\theta_{\rm a}$ does not
become uniform; it tends to the static solution
 \begin{equation}
\lim_{t\to\infty}\theta_{\rm t} = \theta_0\epsilon(z)(1-e^{-e\eta|z|}).
 \end{equation}
Similarly,
 \begin{equation}
\lim_{t\to\infty} A_{\rm t}^z = \eta\theta_0 e^{-e\eta|z|}.
 \end{equation}
In a sense phase equilibration {\it never\/} occurs in this gauge.  Of
course, $D_z\theta_{\rm t}$ does vanish.

Clearly we have to be careful if we use the temporal gauge to allow
for these extra terms in the equations of motion.  Note that the time
derivatives of $\theta_{\rm t}$ and $A_{\rm t}^z$ do obey the
Klein-Gordon equation.

It is also interesting to examine the form of the gauge fields.  We
easily find that the only non-vanishing components of $F^{\mu\nu}$ are
 \begin{equation}
F^{\alpha z} = x^\alpha\partial_z a(\tau,z).
 \end{equation}
In other words, we have a longitudinal electric field,
$E^z=-t\partial_z a$ and azimuthal magnetic field
$B^\phi=\rho\partial_z a$ (in cylindrical polars).

\section{Equations with dissipation}

We now add an extra term to Maxwell's equations:
 \begin{equation}
\partial_\lambda F^{\lambda\mu}=j^\mu+j^\mu_{\rm c},
\label{MaxEq}
 \end{equation}
where $j^\mu_{\rm c}$ is the conduction current.  This represents
the effect of dissipation due to coupling with other fields.  We shall
assume that it can be written in the standard ohmic form, {\it i.e.},
that the spatial part of $j^\mu_{\rm c}$ is given by
 \begin{equation}
{\bf j}_{\rm c}=\sigma{\bf E},
 \end{equation}
for some suitably defined conductivity $\sigma$.  The
corresponding charge density $\rho_{\rm c}$ is fixed by the
requirement that the continuity equation,
 \begin{equation}
\dot\rho_{\rm c} + \nabla\cdot{\bf j}_{\rm c} = 0,
 \end{equation}
hold (together with the intial condition $\rho_{\rm c}=0$ at early
times).  In phase transitions with weak supercooling, bubbles expand
on a background of a dense relativistic plasma which is highly
conductive.  Even in the case of strong supercooling, the energy
released in the collision region creates a plasma of its own, and again
the conductivity is expected to be high.  We shall see that phase
equilibration between colliding bubbles is strongly affected by plasma
dissipation.

Setting $A^\mu=(\phi,{\bf A})$, we may write the equations
(\ref{MaxEq}) as
 \begin{equation}
-\nabla^2\phi-\nabla\cdot\dot{\bf A} =
 -eX^2(\dot\theta+e\phi)+\rho_{\rm c}
 \end{equation}
and
 \begin{equation}
\ddot{\bf A}-\nabla^2{\bf A}+\nabla\nabla\cdot{\bf A}+\nabla\dot\phi =
eX^2(\nabla\theta-e{\bf A})-\sigma(\dot{\bf A}+\nabla\phi).
\label{EqforA}
 \end{equation}
The first of these only serves to define $\rho_{\rm c}$ and is
therefore not very useful.

We should really also include a damping term in the equation for $X$.
However, we shall not consider that explicitly.  Instead, we assume
that the oscillations in the radial direction in field space are
efficiently damped so that when the bubbles collide $X$ rapidly settles
down to its equilibrium value, $X=\eta$.

Let us then specialize to the region inside the bubbles where
$X=\eta$.  In this case, it seems best to consider first the unitary
gauge.  (For the moment we drop the subscript u.)  The equations then
become
 \begin{equation}
\dot\phi+\nabla\cdot{\bf A}=0
 \end{equation}
and
 \begin{equation}
\ddot{\bf A}-\nabla^2{\bf A}+\sigma\dot{\bf A}
+e^2\eta^2{\bf A} = -\sigma\nabla\phi.
 \end{equation}
If the fields vanish outside a finite volume, we can
unambiguously separate the longitudinal and transverse parts of ${\bf
A}$, writing
 \begin{equation}
{\bf A}={\bf A}^{\rm T}+\nabla{1\over\nabla^2}\nabla\cdot{\bf A}.
 \end{equation}
The transverse part satisfies the damped Klein-Gordon equation
with mass $e\eta$:
 \begin{equation}
\ddot{\bf A}^{\rm T}-\nabla^2{\bf A}^{\rm T}
+\sigma\dot{\bf A}^{\rm T}+e^2\eta^2{\bf A}^{\rm T}={\bf 0}.
\label{EoMTrs}
 \end{equation}
If we look for modes proportional to
 \begin{equation}
e^{pt+i{\bf k}\cdot{\bf x}},
 \end{equation}
we find for $p$ the dispersion equation
 \begin{equation}
p = -{\sigma\over2}\pm i\left({\bf k}^2
+e^2\eta^2-{\sigma^2\over4}\right)^{1/2},
\label{EqforP}
 \end{equation}
representing a damped oscillation.

For the longitudinal modes, we find the third-order equation
 \begin{equation}
\overdots{\phi}
-\nabla^2\dot\phi+\sigma\ddot\phi
+e^2\eta^2\dot\phi-\sigma\nabla^2\phi=0,
\label{EoMphi}
 \end{equation}
which yields
 \begin{equation}
(p^2+{\bf k}^2+e^2\eta^2)p+\sigma(p^2+{\bf k}^2)=0.
 \end{equation}

If $\sigma$ is small, there is one real root,
 \begin{equation}
p=p_1\approx-{\sigma{\bf k}^2\over\omega^2},
\label{EigO}
 \end{equation}
where
 \begin{equation}
\omega^2={\bf k}^2+e^2\eta^2,
 \end{equation}
and a complex conjugate pair,
 \begin{equation}
p=p_\pm\approx\pm i\omega-{\sigma e^2\eta^2\over2\omega^2}.
\label{Eigs}
 \end{equation}
For small ${\bf k}$, the latter two modes behave exactly like the
transverse modes, but the first one is a much more slowly decaying
mode with no oscillation.

On the other hand, if $\sigma\gg e\eta$ and $k \ll e^2\eta^2 /\sigma$,
then all three roots are real:
 \begin{equation}
p_1 \approx -{\sigma k^2 \over{e^2\eta^2}},
\qquad p_2 \approx -{e^2\eta^2
\over{\sigma}} + {\sigma k^2\over e^2\eta^2},\qquad
p_3 \approx -\sigma + {e^2\eta^2\over\sigma}.
 \end{equation}
It should be noted that the slowly-decaying mode $p_1$ becomes pure
gauge in the limit $k \to 0$.  In the next section we shall see that
this mode does not affect the evolution of the gauge-invariant phase
difference $\Delta\theta$.

Now let us briefly consider the equations in the temporal gauge.
With $\phi=0$, the equations become
 \begin{equation}
\ddot\theta_{\rm t}-\nabla^2\theta_{\rm t
}-e\nabla\cdot{\bf A}_{\rm t}=0
\label{EoMTheta}
 \end{equation}
and
 \begin{equation}
\ddot{\bf A}_{\rm t}-\nabla^2{\bf A}_{\rm t}
+\nabla\nabla\cdot{\bf A}_{\rm t}
+\sigma\dot{\bf A}_{\rm t}+e^2\eta^2{\bf A}_{\rm t} =
-e\eta^2\nabla\theta_{\rm t}.
 \end{equation}
Of course the equation for the transverse part of ${\bf A}_{\rm t}$ is
exactly the same as before, eq.\ (\ref{EoMTrs}), but this time if we
eliminate $\nabla\cdot{\bf A}_{\rm t}$ using (\ref{EoMTheta}) we find
for $\theta_{\rm t}$ a {\it fourth}-order equation: in fact
$\dot\theta_{\rm t}$ satisfies the same equation (\ref{EoMphi}) as does
$\phi_{\rm u}$.  This is actually obvious because the gauge
transformation between the two gauges shows that $\dot\theta_{\rm t} =
e\phi_{\rm u}$.

In the temporal gauge we always have the freedom of making an arbitrary
time-indenpendent gauge transformation.  Thus the equations have static
solutions in which $\theta$ is {\it any\/} arbitrary function; of
course ${\bf A}$ must be chosen to make $\nabla\theta-e{\bf A}={\bf
0}$.  Note that since only $\dot\theta$ appears in the equation, the
extra terms discussed in the preceding section do not appear; they
would reappear, however, if we tried to write an equation for $\theta$
itself, incorporating the chosen initial conditions.  In a non-simply
connected region, the $\theta_{\rm t}$ equation allows more general
types of solution, in which it is not single-valued.  For example, we
could mimic a closed loop by imposing periodic boundary conditions, and
look for solutions with $\theta(z+L)=\theta(z)+2n\pi$.  One such is
$\theta=2n\pi z/L$, with $A^z=2n\pi/eL$.  This is similar to the type
of  solution one would expect to see eventually around a string, once
the oscillations have died down.

\section{Collisions between planar walls}

As in Section II, we shall first consider the central region of a
two-bubble  collision, where the problem reduces to one-dimensional
form.

In one dimension of course the only modes are longitudinal, so we
should use (\ref{EoMphi}) or its equivalent in the temporal gauge.  We
have assumed that $X$ settles down rapidly to its equilibrium value.
It is reasonable then to take as our initial condition, just after the
bubbles have collided, say at $z=0$ at $t=0$, a configuration where
$X=\eta$ and (in the temporal or axial gauge) $\theta$ has a
discontinuous jump from $-\theta_0$ to $\theta_0$, as in
(\ref{ThetaInit}).  We also assume that at that time there are no
electromagnetic fields present, so we may take
 \begin{equation}
{\bf A}_{\rm t}(0,z) = {\bf 0}, \qquad \dot{\bf A}_{\rm t}(0,z) = 0.
\label{AInit}
 \end{equation}

We can now transform back to the unitary gauge, where we find as
before the initial conditions
 \begin{equation}
\phi(0,z)=0, \qquad \dot\phi(0,z) = {2\theta_0\over e}\delta'(z),
\label{PhiInit}
 \end{equation}
 \begin{equation}
A^z(0,z)= - {2\theta_0\over e}\delta(z), \qquad \dot A^z=0.
\label{bc}
 \end{equation}
The solution is immediate:
 \begin{equation}
\phi(t,z) = \int{dk\over2\pi}e^{ikz}\left(\varphi_1e^{p_1t}
+\varphi_+e^{p_+t}+\varphi_-e^{p_-t}\right),
\label{PhiSol}
 \end{equation}
where $p_1$ and $p_\pm$ are given by (\ref{EigO}) and
(\ref{Eigs}) and the amplitudes $\varphi_1, \varphi_\pm$ satisfy
 \begin{eqnarray}
\varphi_1 + \varphi_+ + \varphi_- =& 0,\nonumber\\
p_1\varphi_1 + p_+\varphi_+ +p_-\varphi_- =& 2ik\theta_0/e,
\label{CoeffEqs}\\
p_1^2\varphi_1 + p_+^2\varphi_+ + p_-^2\varphi_- =&0.\nonumber
 \end{eqnarray}
The solution is
 \begin{equation}
\varphi_1=-{2ik\theta_0\over e}{p_++p_-\over(p_+-p_1)(p_--p_1)},
\label{Coeffs}
 \end{equation}
with two similar equations obtained by permutation of
$(1,+,-)$.

The gauge-invariant phase difference (\ref{GIPD}) can be written as
 \begin{equation}
\Delta\theta \equiv \int_{-\infty}^\infty D_z\theta\,dz
= -e\int_{-\infty}^\infty A_{\rm u}^z\,dz.
\label{DelTheta}
 \end{equation}
and can be evaluated using Eq.(\ref{PhiSol}).  Alternatively, we can
obtain an equation for $\Delta\theta (t)$ by integrating
(\ref{EqforA}),
 \begin{equation}
\left( {d^2 \over{dt^2}} +\sigma{d \over{dt}}
+e^2\eta^2 \right) \Delta\theta
=0.
 \end{equation}
The boundary conditions corresponding to (\ref{bc}) are
 \begin{equation}
\Delta\theta (0) =2\theta_0,~~~~~~~ \Delta {\dot \theta}(0)=0,
 \end{equation}
and the solution is
 \begin{equation}
\Delta\theta (t) =A_+ e^{p_+ t} +A_- e^{p_- t},
\label{Deltaoft}
 \end{equation}
where
 \begin{equation}
p_{\pm} =-{\sigma \over{2}} \pm i\left( e^2\eta^2 -{\sigma^2 \over{4}}
\right)^{1/2}
\label{ppm}
 \end{equation}
and
 \begin{equation}
A_{\pm} = \mp {2\theta_0 p_{\mp} \over{p_+ -p_-}}.
\label{Apm}
 \end{equation}
Note that the values of $p_{\pm}$ are the same as in Eq.(\ref{EqforP})
for the transverse modes at $k=0$.

In the small-$\sigma$ case, we find
  \begin{equation}
\Delta\theta (t) = 2\theta_0 e^{-\sigma t/2}
\left(\cos e\eta t+{\sigma\over 2e\eta}\sin e\eta t\right).
 \end{equation}
It is interesting to compare this with the corresponding expression,
(\ref{ThetaAsy}), in the undamped case.  The oscillating factor is
essentially the same in both cases.  The difference is that here the
dissipation causes exponential damping, while in the previous case the
spherical geometry led to a linear damping.  For large $\sigma$ there
is no oscillation, and Eqs.(\ref{Deltaoft})-(\ref{Apm}) give
 \begin{equation}
\Delta\theta (t) =2\theta_0 \exp (-e^2 \eta^2 t/\sigma ).
 \end{equation}
The characteristic time scale for the current damping, $t_d$, is given
by
 \begin{equation}
t_d \sim 1/\sigma \quad (\sigma \lesssim e\eta),
\qquad t_d \sim \sigma
/e\eta \quad (\sigma \geq e\eta ).
\label{dampingtime}
 \end{equation}

It is also interesting to look at the solution directly in the
temporal gauge.  There, we have to solve a fourth-order equation, so in
addition to (\ref{ThetaInit}) we need initial conditions for
$\ddot\theta$ and  $\overdots\theta$.
These are given by (\ref{AInit}) together with (\ref{EoMTheta}):
 \begin{equation}
\ddot\theta(0,z)=2\theta_0\delta'(z), \qquad
\overdots\theta(0,z)=0.
 \end{equation}
Thus we find
 \begin{equation}
\theta_{\rm t}(t,z) =
\int{dk\over2\pi}e^{ikz}\left(\vartheta_0+\vartheta_1e^{p_1t}
+\vartheta_+e^{p_+t}+\vartheta_-e^{p_-t}\right).
 \end{equation}
It is easy to verify, however, that in fact $\vartheta_0$ vanishes; for
small $\sigma$, we have
 \begin{equation}
\vartheta_0=0, \qquad
\vartheta_1\approx-{2i\theta_0e^2\eta^2\over k\omega^2}, \qquad
\vartheta_\pm\approx-{ik\theta_0\over\omega^2}
\left(1\mp{3ie^2\eta^2\sigma\over2\omega^3}\right).
 \end{equation}
It is interesting to note that in the limit $k\to0$,
$\vartheta_\pm\to0$ while $\vartheta_1\sim-2i\theta_0/k$.  This shows
that as before
 \begin{equation}
\lim_{z\to\pm\infty} \theta_{\rm t} = \pm\theta_0,
 \end{equation}
although $\theta_{\rm t}(t,z)$ approaches its limit more slowly at
late times.

We can also find a solution corresponding to an idealized
one-dimensional version of a three-bubble collision, by imposing
periodic boundary conditions.  We may represent the three bubbles at
the moment when they meet by the segments $[0,L_1]$, $[L_1,L_2]$ and
$[L_2,L]$ of the $z$ axis.  If we choose the initial  phases to be
$0,\theta_1,\theta_2$, where $\theta_1$, $\theta_2-\theta_1$ and
$2\pi-\theta_2$ all lie between $0$ and $\pi$, then it is natural to
impose the boundary condition $\theta_{\rm t}(L)=\theta_{\rm
t}(0)+2\pi$.  In the unitary gauge, the initial conditions for
$\dot\phi$, as in (\ref{PhiInit}), become
 \begin{equation}
\dot\phi(0,z) = {2\theta_1\over e}\delta'(z-L_1)
+ {2(\theta_2-\theta_1)\over e}\delta'(z-L_2)
+ {2(2\pi-\theta_2)\over e}\delta'(z).
 \end{equation}

The solution is similar to (\ref{PhiSol}) but with the Fourier
integral replaced by a sum over the modes $k_n=2n\pi/L$, and the right
hand side of (\ref{CoeffEqs}) replaced by a sum of three terms with
appropriate phases.  An interesting special case is the one where
$L_1=L/3$, $L_2=2L/3$, while $\theta_1=2\pi/3$ and $\theta_2=4\pi/3$.
Then only modes with $n\equiv0\pmod3$ will contribute.  The
non-vanishing coefficients $\varphi_{0,3n}$ and  $\varphi_{\pm,3n}$ are
of essentially the same form as in (\ref{Coeffs}).  Note that there is
no $n=0$ term.  The boundary condition on $\theta_{\rm t}$ must be
accommodated by adding to the Fourier series an extra constant term,
for example the linear term $2\pi z/L$.

\section{Magnetic flux}

Let us now move away from the immediate neighbourhood of the point of
first contact between the bubbles.  In particular, we want to examine
the magnetic fields present in the region where the collision occurs.

For simplicity, we again suppose that the two colliding bubbles have
equal radii, $R$, at the moment of collision.   The main effect of the
curvature is to delay the collision at points off the line of centres.
If the collision occurs at the origin at time $t=0$, then at a
transverse distance $x$, it will occur (to first order in the
curvature) at time $t=x^2/2Rv$, where $v$ is the velocity of the bubble
walls.  As a first approximation, we may expect that the fields near
$x=y=0$ have the same form as in the planar case, while more generally,
for example,
 \begin{equation}
\theta(t,x,y,z) \approx \theta\left(t-{x^2+y^2\over 2Rv},0,0,z\right).
 \end{equation}
This of course has the effect of introducing transverse gradients into
the fields.  Consequently, $j^x$ and $j^y$ are non-zero and in turn
generate transverse components of ${\bf A}$.  The physical result is
the appearance (as in the undamped case) of an azimuthal magnetic
field.  The magnetic flux actually forms a loop around the region where
the collision is occurring.

We are particularly interested in the total magnetic flux.  It is
possible to calculate that without going into the details, by looking
at the line integral of the vector potential.

Consider a rectangular loop $ABCD$, where (see Figure 1)
 \begin{equation}
A,B=(0,0,\mp Z), \qquad C,D=(X,0,\pm Z).
 \end{equation}
The values of $X$ and $Z$ are chosen so that during the time interval
of interest, all four points are within the bubbles, but the path $CD$
passes outside them.  We have assumed that there are no fields outside
the bubbles.  For simplicity, let us suppose that $\Phi$ remains
slightly non-zero, so that the phase $\theta$ is well defined and
interpolates smoothly between its values inside the bubbles.  (It would
of course make no difference if we assumed some other distribution, so
long as $\Phi$ remains very small, but the argument is simpler in this
case.)  It is easy to see that initially, when the bubbles have just
met,
 \begin{equation}
\oint_{ABCD} D_k\theta\,dx^k=0.
 \end{equation}
In fact, using the definition (\ref{DelTheta}), one sees that
$\Delta\theta_{AB}=\Delta\theta_{DC}=2\theta_0$, while
$\Delta\theta_{BC}=\Delta\theta_{DA}=0$.  On the other hand, after a
time $t\gg t_d$, with $t_d$ given by (\ref{dampingtime}),
dissipation has reduced $\Delta\theta_{AB}$ to zero, while
leaving the other three legs unaltered.  Thus
 \begin{equation}
\oint_{ABCD} D_k\theta\,dx^k=-2\theta_0.
\label{LoopInt}
 \end{equation}
By hypothesis the loop contains no zeros of $\Phi$, so $\theta$ is
single-valued.  Thus it follows at once that
 \begin{equation}
\int_{-Z}^Z dz \int_0^X dx\,B^y = \oint_{ABCD} A_kdx^k
= {2\theta_0\over e}.
 \end{equation}

This is exactly what we should expect.  Note that when three bubbles
meet the three magnetic flux tubes combine.  If there is a total net
phase change of $2\pi$ around the loop, the total flux trapped is
exactly one flux quantum, $2\pi/e$.

It is interesting in passing to see how the result (\ref{LoopInt})
emerges in different gauges.  In the unitary gauge, of course, each
$\Delta\theta$ is given by a line integral of ${\bf A}$; the essential
effect is that along $AB$, $A^z$ is non-zero initially, but zero
finally.  In the temporal gauge, the situation is reversed.  There is
no change in the values of $\theta$ at the four corners.  Initially,
$A^z$ vanishes along $AB$, but after the dissipation its line integral
exactly cancels $\theta_B-\theta_A$.  The most complicated situation
occurs in the axial gauge.  Here we find that finally
$\theta_A=\theta_B=0$, which ensures that $\Delta\theta_{AB}$
vanishes.  However, to keep the line integrals along the transverse
sections $BC$ and $DA$ zero, $A^x$ must acquire a non-zero value to
compensate for the transverse gradient of $\theta$; it is easy to check
that it does.

\section{Flux spreading}

At any particular time $t$ after the initial collision, the magnetic
flux forms a ring.  We can readily estimate its size.  The outer edge
occurs where the bubbles are just meeting, namely at a transverse
distance
 \begin{equation}
x=\sqrt{2Rvt}.
\label{Radius}
 \end{equation}
The time taken for the currents generated to be dissipated at any
given point is $t_d$ given by Eq.(\ref{dampingtime}).
The inner edge of the ring occurs where dissipation
is essentially over, namely at
 \begin{equation}
x=\sqrt{2Rv(t-t_d)}.
 \end{equation}
Note that the ring becomes narrower as it expands.  This is because it
is slowing down.  The speed of expansion is $\sqrt{Rv/2t}$.  If $v=c$,
this is always superluminal (the small-angle approximation breaks down
of course when $t$ becomes of order $R$).  In that regime, one should
not think of a tube of magnetic flux moving outwards, but rather of
magnetic flux being created and destroyed at successively larger
radii.  On the other hand, if $v\ll c$, the speed may eventually be
less than $c$.  At that point, we can no longer disregard the region
outside the bubbles.

While the magnetic field is expelled from bubble interiors (as in the
Meissner effect), it has a tendency to spread at the speed of light in
the false vacuum region outside the bubbles.  If the conductivity due
to particles present in that region can be neglected, then the outer
radius of the magnetic flux ring is $x \sim t$, while the inner radius
is given by Eq.(\ref{Radius}).  At $t > R$ most of the flux will escape
to distances greater than the bubble radii.  If a third bubble arrives
after that, the probability of vortex formation in this three-bubble
collision will be strongly suppressed.

Realistically, however, the case of low bubble velocities corresponds
to weak supercooling, when bubbles expand on a background of relatively
dense plasma, and the conductivity in the exterior region is far from
being negligible.  The magnetic field dynamics in a conducting medium
is described by the equation
 \begin{equation}
{\ddot {\bf B}} - {\bf \nabla}^2{\bf B} + \sigma {\dot {\bf B}} = 0.
 \end{equation}
It is easily seen that the characteristic time of magnetic field
variation on a length scale $L$ is $t \sim \sigma L^2$, and thus the
speed of magnetic flux spreading is
 \begin{equation}
v_B \sim L/t \sim (\sigma L)^{-1} .
 \end{equation}
In the limit of very high conductivity, $v_B \to 0$ (the magnetic flux
is ``frozen'' into the plasma).  If the typical bubble radius at
collision is $\xi$ and the velocity is $v$, then the condition for the
flux spreading to be negligible is
 \begin{equation}
\sigma\xi v \gg 1.
\label{Condition}
 \end{equation}

In order to assess the validity of this condition, we shall need
estimates for $\xi$, $\sigma$ and $v$.  The radius $\xi$ can be
estimated as $\xi \sim (v/\gamma)^{1/4}$, where $\gamma$ is the bubble
nucleation rate.  To estimate the conductivity $\sigma$, we represent
the drift velocity of charged particles as
 \begin{equation}
v_d \sim p/\epsilon \sim eE\tau /T,
 \end{equation}
where $p$ is the average momentum gained by a particle between
collisions, $E$ is the electric field, $\tau$ is the mean free time,
$T$ is the temperature, and $\epsilon \sim T$ is the average particle
energy.  The mean free time is $\tau \sim (n{\tilde \sigma})^{-1}$,
where ${\tilde \sigma} \sim e^4 /T^2$ is the characteristic scattering
cross-section and $n$ is the density of charged particles.  Now, the
current density is $j \sim nev_d \sim (T/e^2)E$, and thus
 \begin{equation}
\sigma \sim T/e^2 ,
\label{conductivity}
 \end{equation}
and the condition (\ref{Condition}) can be rewritten as
 \begin{equation}
v \gg e^2 /\xi T .
\label{NewCondition}
 \end{equation}
Since the average bubble separation $\xi$ is expected to be greater
than the thermal particle wavelength $T^{-1}$, the right-hand side of
(\ref{NewCondition}) does not exceed $10^{-2}$.  The temperature at the phase
transition is typically $T \sim \eta$, and Eq.(\ref{conductivity})
corresponds to the high-conductivity regime, $\sigma \gg e\eta$.

The bubble velocity $v$ in the overdamped regime can be roughly
estimated from $f \sim \Delta \rho_v$, where $f$ is the force of
friction due to particle scattering off the bubble wall, and $\Delta
\rho_v$ is the vacuum energy difference between the two sides of the
wall.  Both $f$ and $\Delta \rho_v$ are model-dependent, but in all
cases $f \lesssim T^4 v$, which gives a lower bound on $v$,
 \begin{equation}
v \gtrsim \Delta \rho_v /T^4 .
 \end{equation}
Typically, one finds that bubble velocities are rather high.  For
example, a detailed analysis for the case of electroweak phase
transition gives $v \sim 0.1 - 1$.  This suggests that the condition
(\ref{NewCondition}) is typically satisfied, so that the magnetic flux
does not escape.  Exceptions may occur in cases of nearly degenerate
vacua, when $\Delta \rho_v$ is very small.

\section{Conclusions}

Our main conclusions can be
formulated in terms of the gauge-invariant phase difference $\Delta
\theta$ defined in Eq.(\ref{GIPD}).  We found that $\Delta\theta$
remains equal to its initial value while the bubbles are well-separated
and undergoes damped oscillations after the collision.  The oscillation
period is determined by the gauge boson mass, and the damping time by
the plasma conductivity.  Both timescales are typically much smaller
than the bubble radii at the time of collision.  This is to be
contrasted to the case of a global symmetry breaking, when the phases
in the bubble interiors can change only after the arrival of the
Goldstone boson waves propagating at the speed of light from the
collision region.

The currents associated with the oscillating phase gradients
generate a magnetic field which is concentrated in a ring-shaped region
near the circle on which the bubble walls intersect.  The total
magnetic flux is determined by the initial value of $\Delta\theta$,
 \begin{equation}
\Phi_B = \Delta\theta_0 /e.
 \end{equation}
Right after the collision, the radius of the intersection circle grows
faster than the speed of light, but if the bubble expansion is
substantially slowed down by interaction with ambient plasma, the speed
may eventually become nonrelativistic.  The magnetic field then begins
to spread into the exterior region and can in principle escape to
distances large compared to the bubble radii.  However, magnetic field
has a tendency to be ``frozen'' into a conducting plasma, and the speed
of spreading is typically smaller than the bubble expansion speed (with
the exception of cases of nearly degenerate vacua, when the bubble
velocities are extremely small).

In a three-bubble collision, the three magnetic flux tubes combine
when the bubbles meet.  The resulting magnetic flux is $\Phi_B =
\Delta\theta_{\rm tot} /e$, where $\Delta\theta_{\rm tot} = 2\pi n$ is
the sum of the initial phase differences between the bubbles and $n$ is
an integer.  For $n \ne 0$, a vortex is formed carrying $n$ flux
quanta, $\Phi_B = 2\pi n/e$.  Thus, although the physics of bubble
collisions turned out to be interesting and complicated, our
conclusions are essentially the same as those resulting from the naive
analysis in Ref.\ \cite{Kib76}.

In conclusion, we would like to mention some open questions.  As a
matter of principle, it would be interesting to understand what happens
when the bubble expansion speed is so low that most of the magnetic
flux does escape.  One expects that the vortex density in this case is
suppressed, and we are currently investigating what effect this may
have on the properties of the resulting string distriution.

Throughout the paper we assumed that the radial mode of the Higgs
field is strongly damped and settles to its equilibrium value on a
timescale short compared to the phase equilibration process.  In
general, however, the energy dissipation in the radial mode is itself a
complicated process and can take considerable time, especially in the
case when the colliding walls are relativistic.  The phase
equilibration in this case can be studied using numerical simulations.

Another interesting and important question is what effect thermal
fluctuations of the magnetic field and of winding number have on the
vortex formation.  A naive estimate of the magnetic flux fluctuation on
scale $\xi$ is $\delta\Phi_B \sim (\xi T)^{1/2}$, where $T$ is the
temperature.  This could in principle be much greater than a single
flux quantum.  If all this flux is squeezed into a vortex by advancing
bubble walls, then we can expect that a sizable proportion of vortices
will have large winding numbers.  A particularly interesting situation
may arise in models where the gauge bosons are heavier than the Higgs
particles, so that $n>1$ strings are stable.  In this case one expects
the formation of an interconnected network of strings with different
winding numbers.  The cosmological evolution of such a network may be
very different from the standard picture.

Thermal fluctuations may also play an important role in defect
formation during first-order phase transitions with a global symmetry
breaking.  In this case, the phase $\theta$ is a massless Goldstone
field, and its fluctuation on scale $\xi$ is $\delta\theta \sim
(T\xi)^{1/2}$.  Once again, this can be large, resulting in the
formation of large vortex clusters.  We note, however, that  multiply
charged vortices are unstable in the case of a global symmetry
breaking, and we expect the clusters to disperse.

\acknowledgments
We are indebted for hospitality to the Isaac Newton Institute
for Mathematical Sciences, Cambridge, England, where this work was
begun.  We have benefited from discussions with numerous participants,
in particular Allen Everett, Eugene Chudnovsky, Mark Hindmarsh, Tanmay
Vachaspati and Julian Borrill.

\begin{figure}

\caption{}{Two colliding bubbles, showing the closed path $ABCD$.}

\end{figure}

 \end{document}